\documentclass[preprint,12pt]{elsarticle}

\usepackage{amssymb}
\usepackage{amsmath}
\usepackage{float}
\usepackage{booktabs}
\usepackage{array}
\usepackage{longtable}
\usepackage{caption}
\usepackage{threeparttable}
\usepackage{geometry}
\usepackage{url}
\usepackage{rotating}
\usepackage{graphicx}
\usepackage{subcaption}
\usepackage{placeins}

\begin{document}

\begin{frontmatter}



\title{Urban Structure Predicts a Persistent Place-Based Spatial Signature of Dengue Susceptibility Across Epidemic Conditions}


\author[1]{Marcílio Ferreira dos Santos\corref{cor1}}
\ead{marcilio.santos@ufpe.br}
\ead[url]{https://orcid.org/0000-0001-8724-0899}

\author[2]{Cleiton de Lima Ricardo}
\ead{cleiton.lricardo@ufpe.br}
\ead[url]{https://orcid.org/0000-0002-7114-1201}

\author[3]{Andreza dos Santos Rodrigues de Melo}
\ead{andreza.rodrigues@ufpe.br}
\ead[url]{https://orcid.org/0000-0002-5153-6548}

\author[1]{José Dilson Beserra Cavalcanti}
\ead{dilson.cavalcanti@ufpe.br}
\ead[url]{https://orcid.org/0000-0002-6125-3867}

\author[4]{Larissa Suellen Gomes Andrade de Lima Aquino}
\ead[url]{https://orcid.org/0009-0003-8960-3157}

\cortext[cor1]{Corresponding author.}


\affiliation[1]{
    organization={Núcleo de Formação de Docentes, Universidade Federal de Pernambuco (UFPE)},
    city={Caruaru},
    state={PE},
    country={Brazil}
}

\affiliation[2]{
    organization={Núcleo Interdisciplinar de Ciências Exatas e da Natureza (NICEN), Universidade Federal de Pernambuco (UFPE)},
    city={Caruaru},
    state={PE},
    country={Brazil}
}

\affiliation[3]{
    organization={Departamento de Engenharia Cartográfica, Universidade Federal de Pernambuco (UFPE)},
    city={Recife},
    state={PE},
    country={Brazil}
}

\affiliation[4]{
    organization={Secretaria de Educação de Pernambuco},
    state={Pernambuco},
    country={Brazil}
}

\begin{abstract}

Urban structure alone consistently recovered the relative spatial distribution
of dengue susceptibility in unseen years, with year-specific Spearman
correlations ranging from 0.49 to 0.76. Structural predictors achieved stronger
mean prospective spatial ranking than recent epidemiological history, while
combining both information sources produced little additional gain. Temporal
generalization collapsed when year-specific spatial correspondence was
disrupted (null mean Spearman = 0.0048; empirical $p=0.002$), but was
reproduced when coherent place-level susceptibility trajectories were
preserved. Cross-classification models retained 93.79\% of within-classification
ranking performance after matching training-set size, although the corresponding
transportability gap was not unusual relative to random partitions of epidemic
years ($p=0.8024$). These findings support a persistent place-based spatial
hierarchy of dengue susceptibility that remains informative across markedly
different epidemic conditions. Measured urban structure provides an informative
predictive representation of this hierarchy, but does not constitute a complete
or causal explanation of its persistence. Persistent susceptibility mapping may
therefore complement short-term epidemiological surveillance by supporting
longer-term, place-based urban health planning.

\end{abstract}
\begin{graphicalabstract}
\end{graphicalabstract}

\begin{highlights}
\item Urban structure predicts a persistent place-based hierarchy of dengue susceptibility
\item Structural models generalize across ten years with Spearman correlations of 0.49--0.76
\item Temporal predictability collapses when persistent spatial correspondence is disrupted
\item Urban structure outperforms recent epidemiological history in mean spatial risk ranking
\item Susceptibility remains broadly transportable across contrasting epidemic intensities
\end{highlights}

\end{frontmatter}



\section{Introduction}

Dengue remains one of the most important mosquito-borne diseases affecting
tropical and subtropical cities, where recurrent epidemics impose substantial
and spatially uneven public-health burdens. Its distribution within cities is
rarely homogeneous: neighborhoods may experience markedly different levels of
incidence according to population density, socioeconomic conditions, housing
and sanitation, vegetation, drainage infrastructure, and other characteristics
of the built and social environment
\cite{costa1998dengue,flauzino2009heterogeneidade,andrioli2020spatial,
urbanDengueReview2022}. Such heterogeneity is especially relevant in rapidly
urbanizing and socially unequal settings, where environmental exposure,
infrastructure, and socioeconomic vulnerability interact at fine spatial
scales. From an urban sustainability perspective, understanding these
persistent differences is important because climate-sensitive health risks
affect not only disease surveillance but also urban resilience, infrastructure
planning, and the capacity of cities to anticipate and respond to recurrent
epidemics \cite{sahay2018urban,sahay2019adaptation}.

A growing body of evidence suggests that dengue risk is partly embedded in the
organization of urban space. Spatial epidemiological studies have associated
dengue incidence with socioeconomic deprivation, population characteristics,
land cover, environmental conditions, and characteristics of the built
environment
\cite{costa1998dengue,flauzino2009heterogeneidade,andrioli2020spatial,
urbanDengueReview2022}. Recent evidence from Recife further indicates that the
functional organization of the city may contain epidemiologically relevant
information beyond conventional geographic proximity. In particular,
network-mediated connectivity derived from the urban road system was found to
explain spatial variation in dengue incidence across multiple statistical and
spatial modeling frameworks \cite{santos2026curvature}. These findings suggest
that dengue risk may reflect relatively persistent properties of urban places,
rather than only short-term epidemic fluctuations.

Most dengue prediction research, however, has focused on a different problem:
forecasting future incidence or outbreak intensity from climatic conditions,
recent epidemiological history, environmental variables, and spatiotemporal or
machine-learning models
\cite{salles2018history,zheng2019spatiotemporal,hoyos2021machine,
roster2022machine}. These approaches are valuable for anticipating short-term
changes in transmission, and recent case history can provide substantial
predictive information for forecasting dengue incidence
\cite{roster2022machine}. Comparatively less attention has been given to a
different question: whether the \emph{relative spatial distribution of dengue
susceptibility within a city is itself persistent across years}. Citywide
incidence may change dramatically between epidemic seasons, whereas many
components of the urban environment associated with spatial
heterogeneity---including built form, population distribution, socioeconomic
conditions, and infrastructure---evolve on considerably slower timescales.

This distinction between \emph{epidemic magnitude} and \emph{spatial
susceptibility} has important implications for urban health planning. If the
relative distribution of neighborhood risk is dominated by recent transmission
history, spatial priorities may need to be substantially reconstructed after
each epidemic cycle. Conversely, if relatively stable characteristics of urban
places encode a persistent spatial signature of susceptibility, some areas may
remain systematically more vulnerable even when the overall magnitude of
dengue changes. Structural risk mapping could then complement short-term
epidemic forecasting by supporting sustained surveillance, vector-control
prioritization, and longer-term interventions in vulnerable urban areas
\cite{sahay2018urban,sahay2019adaptation}. Testing this hypothesis requires
temporally explicit out-of-sample validation capable of distinguishing
persistent place-based structure from recent epidemiological history and
period-specific predictability.

Recife, Brazil, provides an informative setting in which to investigate this
problem. The city combines pronounced intra-urban socioeconomic and
environmental heterogeneity with large interannual fluctuations in dengue
incidence, while previous research has demonstrated substantial spatial
variation in dengue vulnerability and epidemiologically meaningful structure
in its urban environment
\cite{andrioli2020spatial,santos2026curvature}. Using 95,665 reported dengue
cases distributed across 93 neighborhoods from 2015 to 2024, together with
seven neighborhood-level indicators describing population density,
socioeconomic conditions, informal urbanization, vegetation, drainage
infrastructure, household occupancy, and housing vacancy, we examine whether
relatively stable urban characteristics recover a persistent signal of
relative dengue susceptibility. Rather than focusing primarily on absolute case
prediction, we separate the relative spatial distribution of susceptibility
from changes in citywide epidemic magnitude.

Specifically, we compare models based exclusively on urban structural
characteristics with models based on recent epidemiological history and with
models combining both information sources. We evaluate generalization using
rolling-origin prospective and leave-one-year-out validation, followed by
cross-intensity transfer experiments and null-model analyses designed to
distinguish persistent place-based organization from epidemic-specific
predictability. Additional spatial and model-class robustness analyses assess
whether the resulting pattern can be explained by persistent residual spatial
dependence or by the choice of predictive algorithm.

We hypothesize that dengue susceptibility exhibits a persistent and
transferable place-based spatial signature that can be partially recovered
from relatively stable characteristics of the urban environment and cannot be
reduced to recent epidemiological history alone. We do not assume that the
measured structural indicators constitute the unique or causal determinants of
this persistence; rather, we test whether they provide an observable
representation of a broader and temporally stable spatial organization of
susceptibility. By shifting attention from predicting only \emph{how large} a
future epidemic may become toward identifying \emph{where} its burden is
consistently more likely to concentrate, this study connects short-term dengue
forecasting with longer-term urban vulnerability assessment and provides a
framework for anticipatory risk mapping and geographically targeted health
planning.

\section{Materials and Methods}

\subsection{Study Area, Data Sources, and Panel Construction}

This study was conducted in Recife, Pernambuco, northeastern Brazil, a coastal
city with approximately 1.49 million inhabitants, 219 km$^2$, and 93
neighborhoods \cite{ibge2024}.

Dengue surveillance records for 2015--2024 were obtained from the municipal
open-data infrastructure and aggregated by neighborhood and month. Duplicate
neighborhood--month records were removed without summing duplicated counts.
After cleaning, the dataset contained 95,665 dengue cases.

Seven neighborhood-level structural indicators were obtained primarily from
the 2022 Brazilian Demographic Census (IBGE): population density, proportion
of informal settlements, vegetation proportion, channel proportion, average
household occupancy, vacant household rate, and collective household rate.
These variables were time-invariant within neighborhoods and were treated as
comparatively persistent spatial attributes throughout the study period.

Neighborhood--month combinations without case records were interpreted as zero
recorded cases rather than missing observations. Reconstruction of the complete
panel yielded 11,160 observations (93 neighborhoods $\times$ 120 months).
Aggregation by year produced the primary analytical panel of 930
neighborhood--year observations (93 neighborhoods $\times$ 10 years), preserving
the total of 95,665 cases.

\subsection{Outcome, Predictors, and Modeling Framework}

To separate relative spatial susceptibility from annual epidemic magnitude,
neighborhood relative risk was defined as

$$
RR_{i,t}
=
\frac{I_{i,t}}{I_{\mathrm{city},t}},
$$

where $I_{i,t}$ and $I_{\mathrm{city},t}$ denote neighborhood and citywide
dengue incidence, respectively. The modeling target was

$$
Y_{i,t}
=
\log(RR_{i,t}+10^{-6}),
$$

with the small constant accommodating neighborhood--year observations with zero
recorded cases. Prediction therefore focused on the within-year spatial
distribution of relative dengue incidence rather than citywide epidemic
magnitude. Throughout the study, ``relative spatial susceptibility'' refers
operationally to this neighborhood-level measure and not to individual
biological susceptibility.

Three predictor specifications were evaluated. The \textit{structure-only}
model used the seven time-invariant urban indicators. The \textit{history-only}
model used seven lagged epidemiological predictors: case counts at one- and
two-year lags, previous-year incidence, relative risk at one- and two-year
lags, and two-year means of case counts and relative risk. The
\textit{combined} model included all 14 predictors. Temporal predictors were
constructed within neighborhoods using only information preceding the
prediction year.

Random Forest regression \cite{breiman2001} was the principal learning
algorithm. All models used 1,000 trees, $\sqrt{p}$ predictors considered at
each split, a minimum terminal-leaf size of 2, a minimum split size of 4, and
bootstrap sampling. This specification was fixed across experiments, random
seeds were controlled, and no hyperparameter optimization used held-out-year
performance.

For prospective comparisons, all model specifications used identical training
observations with complete two-year epidemiological histories. A one-year
persistence benchmark was defined as

$$
\widehat{Y}_{i,t}^{\,\mathrm{pers}}
=
Y_{i,t-1}.
$$

Spearman correlation was the primary performance metric because the principal
objective was to recover the relative spatial ranking of susceptibility.
Pearson correlation, RMSE, MAE, $R^2$, and Precision@10 were used as
complementary metrics. Precision@10 measured the overlap between the ten
highest predicted and observed susceptibility neighborhoods.

\subsection{Temporal Validation and Epidemic-Intensity Transportability}

Temporally explicit out-of-sample validation was used instead of random
train--test splitting \cite{roberts2017}. Rolling-origin validation evaluated
prospective prediction for 2020--2024. For each test year $t$, models were
trained on observations from 2017 through $t-1$ and evaluated on all 93
neighborhoods in year $t$, using only information available before the test
year.

Leave-one-year-out (LOYO) validation evaluated the structure-only model across
2015--2024. Each year was excluded in turn, the model was fitted on the
remaining nine years, and predictions were generated for all neighborhoods in
the held-out year. LOYO therefore evaluates generalization to unseen years for
known spatial units, rather than extrapolation to previously unseen
neighborhoods.

For each year, performance was quantified by the Spearman correlation
$\rho_t$, with overall temporal generalization defined as

$$
\bar{\rho}_{\mathrm{LOYO}}
=
\frac{1}{10}
\sum_{t=2015}^{2024}\rho_t.
$$

Year-specific uncertainty was estimated using 2,000 nonparametric bootstrap
resamples of neighborhoods. These intervals quantify empirical uncertainty in
ranking performance and do not explicitly account for spatial dependence.

To examine whether the spatial susceptibility signal remained informative
across contrasting levels of citywide epidemic activity, years were ranked by
annual dengue incidence and classified descriptively as LOW (2017, 2018,
2022), MID (2019, 2020, 2023), or HIGH (2015, 2016, 2021, 2024). Annual
incidence ranged from approximately 166 to 2,241 cases per 100,000
inhabitants. These groups represent descriptive epidemic-intensity strata and
should not be interpreted as mechanistically distinct epidemic regimes.

Cross-intensity transportability was evaluated by training on years belonging
to intensity groups different from that of the held-out year and comparing
performance with training on years from the same intensity classification.
Because the corresponding training pools differed in size, the larger pool was
randomly subsampled to match the smaller one before model fitting. This
matched-sample procedure was repeated 100 times.

Transportability was summarized by

$$
\Delta_{\mathrm{transport}}
=
\bar{\rho}_{\mathrm{within}}
-
\bar{\rho}_{\mathrm{cross}},
$$

and

$$
R_{\mathrm{transport}}
=
100
\frac{\bar{\rho}_{\mathrm{cross}}}
{\bar{\rho}_{\mathrm{within}}}.
$$

Smaller transportability gaps and higher performance-retention values indicate
greater descriptive stability under cross-intensity transfer. Whether this
behavior was specifically associated with the observed LOW/MID/HIGH
classification was evaluated using permutation-based null models.

\subsection{Statistical Inference, Null Models, and Spatial Sensitivity}

Bootstrap confidence intervals and paired Wilcoxon signed-rank tests were used
for selected comparisons of predictive performance. Random Forest feature
importance and permutation-based feature importance were interpreted only as
measures of predictive relevance and not as estimates of causal effects.

Residual spatial dependence was evaluated using Moran's $I$
\cite{moran1950,anselin1995} and a row-standardized Queen-contiguity spatial
weights matrix. The principal diagnostic used each neighborhood's mean annual
percentile prediction error across LOYO years and 999 permutations. Annual
Moran's $I$ tests were additionally performed on log-relative-risk prediction
errors, with Benjamini--Hochberg correction applied across the ten annual
tests.

Two permutation null models were used to investigate the source of LOYO
predictability. The \textit{independent-year null} independently permuted
neighborhood identities within each year, thereby destroying persistent
cross-year correspondence between place and susceptibility while preserving
the year-specific outcome distributions. The \textit{coherent-place null}
applied a single common neighborhood permutation across all years, preserving
longitudinal place-level susceptibility trajectories while disrupting their
original alignment with the measured structural profiles. Each null model used
500 permutations and repeated the complete LOYO procedure.

For a test statistic $T$, the empirical upper-tail probability was calculated
as

$$
p_{\mathrm{emp}}
=
\frac{
1+\sum_{b=1}^{B}
\mathbb{I}(T_b \geq T_{\mathrm{obs}})
}{
B+1
},
\qquad B=500.
$$

The independent-year and coherent-place nulls were designed to distinguish
predictability associated with persistent place-specific spatial organization
from predictability requiring the original correspondence between observed
susceptibility trajectories and measured structural profiles. They do not
constitute tests of causal effects of the structural covariates.

A complementary null analysis evaluated whether cross-intensity
transportability depended specifically on the observed LOW/MID/HIGH
classification. Intensity labels were randomly reassigned among years while
preserving the observed group sizes (3/3/4). Matched sampling was repeated
20 times within each of 500 permutations. The resulting null distributions
were compared with the observed within-classification and
cross-classification Spearman correlations,
$\Delta_{\mathrm{transport}}$, and $R_{\mathrm{transport}}$.

Finally, sensitivity to Pau Ferro, an atypical neighborhood with a recorded
population denominator of 89 inhabitants and one dengue case during
2015--2024, was evaluated by repeating the principal LOYO and matched
cross-intensity analyses after its exclusion. The reduced dataset contained
92 neighborhoods and 920 neighborhood--year observations.

\subsection{Model-Class Robustness}

To assess whether the principal findings depended on the Random Forest model
class, the structure-only experiments were independently replicated using
gradient-boosted decision trees implemented with XGBoost. The same seven
structural predictors, temporal training--test partitions, and performance
metrics used in the principal Random Forest analyses were retained.

Model-class robustness was evaluated under both rolling-origin prospective
validation (2020--2024) and LOYO validation (2015--2024). XGBoost performance
was compared with the corresponding Random Forest results using year-specific
Spearman correlations, paired bootstrap confidence intervals, and Wilcoxon
signed-rank tests.

Agreement between model classes was additionally quantified using the Spearman
correlation between neighborhood prediction rankings produced by Random Forest
and XGBoost within each LOYO test year. Agreement between their persistent
spatial rankings, obtained by aggregating out-of-sample predictions across
years, was also evaluated.

To assess sensitivity to XGBoost specification without selecting models
according to held-out performance, 24 predefined hyperparameter configurations
were evaluated under the same temporal-validation framework. These experiments
were used exclusively as a robustness analysis and did not replace the
prespecified Random Forest model as the principal modeling framework.

All analyses were implemented in Python using \texttt{pandas}, \texttt{NumPy},
\texttt{SciPy}, \texttt{scikit-learn}, \texttt{xgboost}, and
\texttt{esda}/PySAL, with fixed random seeds. Detailed model settings,
year-specific results, bootstrap and permutation procedures,
feature-importance analyses, spatial diagnostics, model-class robustness
analyses, and sensitivity results are provided in the Supplementary Material.
\section{Results}

\subsection{Dengue burden and prospective prediction}

The final analytical dataset comprised 95,665 dengue cases distributed across
93 neighborhoods and ten epidemiological years (2015--2024), resulting in a
balanced panel of 930 neighborhood--year observations. Annual dengue activity
varied markedly throughout the study period, providing contrasting
epidemiological conditions under which to evaluate the persistence of
neighborhood-level susceptibility.

Citywide incidence ranged from approximately 166 to 2,241 cases per 100,000
inhabitants. The highest epidemic activity occurred in 2015
(2,241.2 cases per 100,000 inhabitants), followed by 2016
(1,286.0 per 100,000), whereas the lowest incidence was observed in 2017
(165.5 per 100,000).

Following the rank-based classification defined in the Methods, the three
years with the lowest citywide incidence (2017, 2018, and 2022) were assigned
to the low-incidence group, the next three (2019, 2020, and 2023) to the
intermediate-incidence group, and the four years with the highest incidence
(2015, 2016, 2021, and 2024) to the high-incidence group. The classification
therefore spanned more than an order of magnitude in annual epidemic intensity.


\begin{table}[htbp]
\centering
\caption{Annual dengue burden and epidemic-intensity classification in Recife,
Brazil, 2015--2024. Years were ranked according to annual citywide dengue
incidence, with the three lowest-incidence years classified as low, the next
three as intermediate, and the four highest-incidence years as high.}
\label{tab:annual_regimes}

\begin{tabular}{rrrr}
\hline
Year & Dengue cases & Incidence per 100,000 & Epidemic intensity \\
\hline
2015 & 33,190 & 2241.168 & High \\
2016 & 19,044 & 1285.954 & High \\
2017 &  2,451 &  165.505 & Low \\
2018 &  2,624 &  177.187 & Low \\
2019 &  7,260 &  490.234 & Intermediate \\
2020 &  3,526 &  238.095 & Intermediate \\
2021 & 10,818 &  730.490 & High \\
2022 &  2,873 &  194.001 & Low \\
2023 &  3,493 &  235.866 & Intermediate \\
2024 & 10,386 &  701.319 & High \\
\hline
\end{tabular}

\end{table}

Against this heterogeneous epidemiological background, we first examined
whether persistent urban characteristics contained prospective predictive
information on the relative spatial distribution of dengue susceptibility and
how this information compared with recent epidemiological history. Under
rolling-origin validation for 2020--2024, the structure-only model achieved
the highest mean Spearman correlation ($\bar{\rho}=0.5913$), compared with
0.5165 for the history-only model, 0.5463 for the combined model, and 0.5452
for the one-year persistence benchmark.

The structure-only model also showed the lowest mean RMSE among the three
Random Forest specifications (1.3678), followed by the combined (1.4126)
and history-only (1.4443) models. Mean absolute errors were 0.6071, 0.6424,
and 0.6642, respectively. Mean $R^2$ values were 0.5259 for the
structure-only model, 0.4929 for the combined model, and 0.4577 for the
history-only model. The persistence benchmark showed substantially poorer
magnitude-based performance ($\mathrm{RMSE}=1.8416$; mean $R^2=0.0170$),
despite a mean Spearman correlation similar to that of the combined model.

Paired bootstrap comparisons supported an advantage of structural information
over recent epidemiological history. The structure-only model exceeded the
history-only model by a mean Spearman difference of
$\Delta\rho=0.0748$ (95\% bootstrap CI: 0.0153--0.1318). The corresponding
year-level Wilcoxon signed-rank comparison was not statistically significant
($p=0.125$), consistent with the limited number of prospective test years
($n=5$).

Combining structural and epidemiological-history predictors did not improve
prospective ranking relative to the structure-only specification. The mean
difference between the combined and structure-only models was
$\Delta\rho=-0.0450$ (95\% bootstrap CI: $-0.0960$ to $0.0026$).
Similarly, the advantage of the combined model over the history-only model was
smaller and its confidence interval included zero
($\Delta\rho=0.0298$, 95\% bootstrap CI: $-0.0009$ to $0.0597$).

These results indicate that the seven persistent structural characteristics
contained most of the prospective predictive information recovered by the
evaluated predictor sets, with no consistent additional gain from incorporating
recent epidemiological history. They establish prospective predictive value
for the measured structural variables, while the extent to which temporal
persistence can be specifically attributed to them is addressed by the
null-model analyses below.


\begin{table}[htbp]
\centering
\caption{Mean out-of-sample predictive performance across the five
rolling-origin prospective test years (2020--2024). Structure-only,
history-only, and combined models were fitted using identical eligible
training observations. The persistence benchmark corresponds to observed
neighborhood susceptibility in the immediately preceding year. Spearman
correlation was the primary performance metric.}
\label{tab:ablation}

\begin{tabular}{lrrrrrr}
\hline
Model & RMSE & MAE & $R^2$ & Pearson & Spearman & Precision@10 \\
\hline
Structure-only    & 1.3678 & 0.6071 & 0.5259 & 0.7205 & 0.5913 & 0.44 \\
History-only      & 1.4443 & 0.6642 & 0.4577 & 0.6959 & 0.5165 & 0.42 \\
Combined          & 1.4126 & 0.6424 & 0.4929 & 0.7098 & 0.5463 & 0.40 \\
Persistence $t-1$ & 1.8416 & 0.6998 & 0.0170 & 0.6164 & 0.5452 & 0.48 \\
\hline
\end{tabular}

\end{table}



\subsection{Structural susceptibility generalizes across epidemic years}

The persistence of the spatial susceptibility signal was subsequently evaluated
using structure-only models under leave-one-year-out (LOYO) validation. In each
iteration, all 93 observations belonging to one epidemiological year were
excluded from model training and used exclusively for testing. Repeating this
procedure across all ten years generated out-of-sample predictions for the full
set of 930 neighborhood--year observations.

The structure-only model retained substantial spatial ranking performance
throughout the study period. Year-specific Spearman correlations were 0.6974 in
2015, 0.6234 in 2016, 0.6667 in 2017, 0.6188 in 2018, 0.7169 in 2019, 0.4907
in 2020, 0.7572 in 2021, 0.5836 in 2022, 0.6290 in 2023, and 0.6095 in 2024.
The strongest correspondence between predicted and observed neighborhood
susceptibility occurred in 2021 ($\rho=0.7572$), whereas the weakest occurred
in 2020 ($\rho=0.4907$). Mean LOYO Spearman correlation across the ten
held-out years was $\bar{\rho}=0.6393$. Ranking performance remained positive
in every held-out year despite substantial variation in citywide dengue
incidence over the decade.

Permutation analyses were used to determine whether this temporal
generalization depended on persistent spatial organization across years. Under
the independent-year null, neighborhood identities were permuted independently
within each year, disrupting cross-year correspondence between place and
susceptibility while preserving the year-specific outcome distribution.
Mean LOYO Spearman correlation collapsed to near zero
(null mean $=0.00481$, SD $=0.04387$; 95\% null interval:
$-0.08171$ to $0.08731$), compared with an observed mean correlation of
$0.63933$. The observed statistic exceeded all 500 independent-year null
realizations (empirical $p=0.0020$; observed percentile $=100\%$).

The complementary coherent-place null preserved the longitudinal
susceptibility trajectory associated with each spatial unit while disrupting
its original correspondence with the measured structural profiles. Under this
null, mean LOYO performance was virtually identical to the observed value
(null mean $=0.63930$, SD $=0.00057$; 95\% null interval:
$0.63818$ to $0.64038$). The observed mean correlation
($\bar{\rho}=0.63933$) occupied the 51.6th percentile of the null distribution
and was not unusually large relative to it (empirical $p=0.4850$).

These complementary null models refine the interpretation of temporal
generalization. The collapse of predictive performance under independent
year-wise permutation indicates that LOYO performance depends strongly on
persistent place-specific spatial organization rather than arbitrary
year-specific associations. At the same time, the near-complete reproduction
of observed performance under the coherent-place null shows that temporal
persistence cannot be uniquely attributed to the seven measured structural
covariates. The results therefore support a persistent place-based spatial
hierarchy of dengue susceptibility, for which the measured urban structural
characteristics provide an informative predictive representation.


\begin{figure}[htbp]
\centering
\includegraphics[width=\textwidth]
{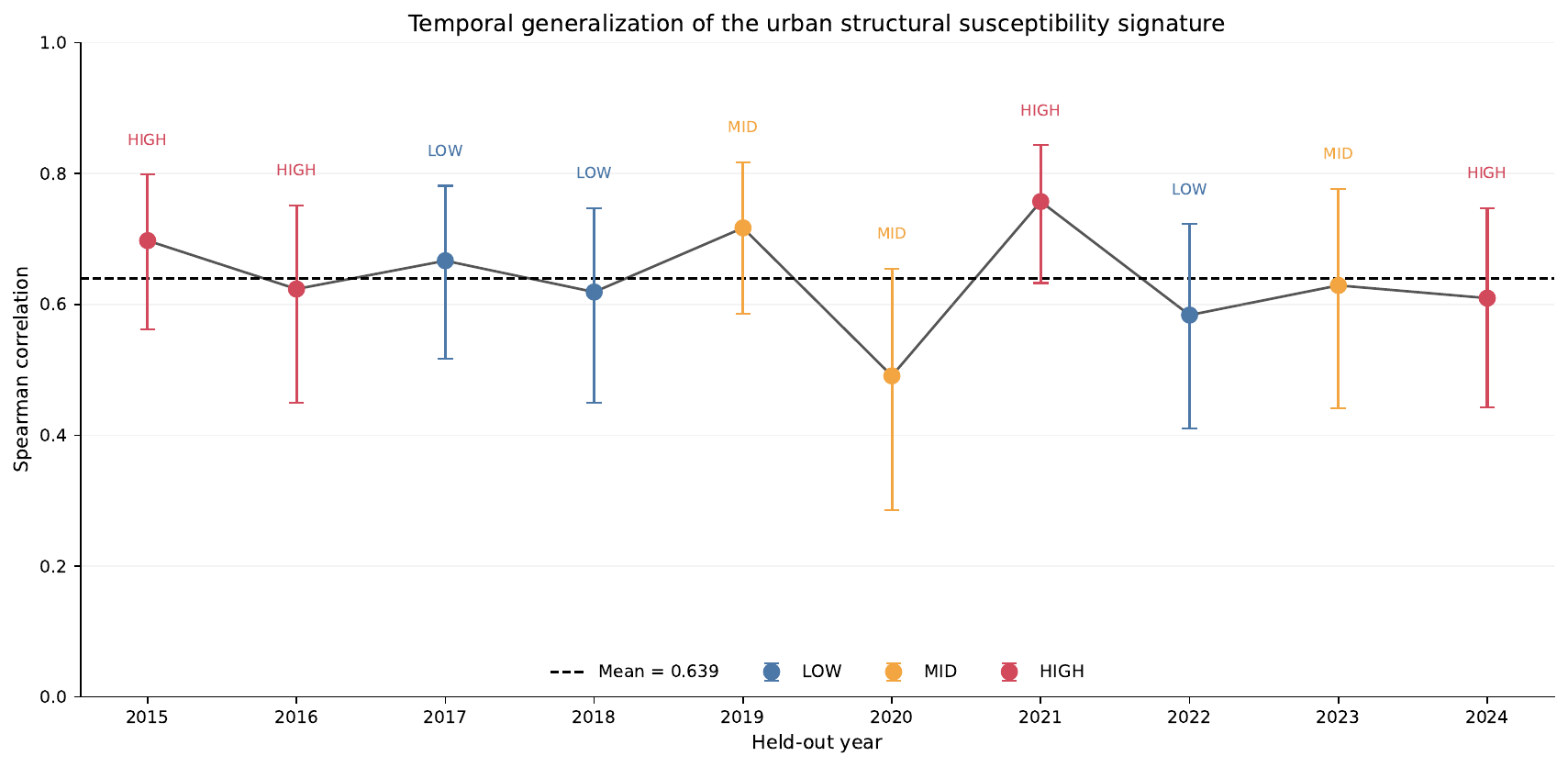}
\caption{Temporal generalization of neighborhood-level dengue susceptibility
across Recife, 2015--2024. Points represent year-specific Spearman
correlations between observed and predicted log-relative risk obtained from
structure-only Random Forest models under leave-one-year-out (LOYO)
validation. Error bars represent 95\% nonparametric bootstrap confidence
intervals. Labels indicate the epidemic-intensity classification of each
held-out year. The dashed horizontal line represents the mean Spearman
correlation across all held-out years ($\bar{\rho}=0.6393$).}
\label{fig:loyo}
\end{figure}



\subsection{The spatial susceptibility signature is transportable across epidemic conditions}

We next examined whether the spatial susceptibility signal remained
recoverable across contrasting levels of epidemic activity. Under LOYO
validation, mean Spearman correlation was 0.672 during high-incidence years
(95\% bootstrap interval: 0.599--0.729), 0.623 during low-incidence years
(95\% bootstrap interval: 0.529--0.698), and 0.612 during
intermediate-incidence years (95\% bootstrap interval: 0.505--0.694).
Predictive ranking therefore remained substantial across all three
epidemic-intensity groups.


\begin{figure}[htbp]
\centering
\includegraphics[width=0.75\textwidth]
{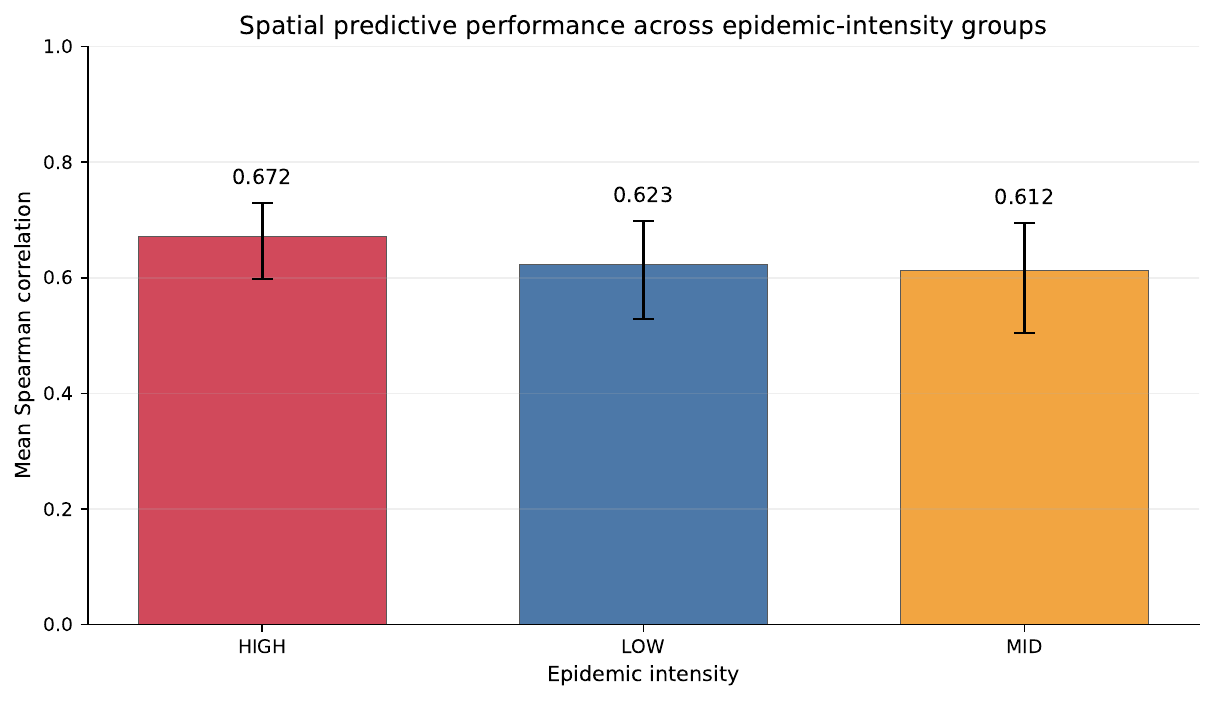}
\caption{Out-of-sample predictive performance of the structure-only model
across epidemic-intensity groups under leave-one-year-out validation. Bars
represent mean Spearman correlations between observed and predicted
neighborhood log-relative risk for held-out years classified as high-, low-,
or intermediate-incidence. Error bars represent 95\% bootstrap intervals
obtained by resampling neighborhoods within held-out years.}
\label{fig:regime_performance}
\end{figure}

Persistence of performance within each intensity group does not necessarily
establish transportability between them. We therefore trained structure-only
models using observations from one epidemic-intensity group and evaluated them
on observations belonging to another. Substantial spatial ranking performance
was preserved across all six transfer directions. Spearman correlations were
0.6361 for low-to-high and 0.6312 for high-to-low transfer. Transfers involving
intermediate-incidence years yielded correlations of 0.5647 for
intermediate-to-high, 0.5279 for intermediate-to-low, 0.6143 for
low-to-intermediate, and 0.5869 for high-to-intermediate transfer.

Pearson correlations ranged from 0.6135 to 0.7567, while cross-intensity
$R^2$ values remained positive in all six experiments
(range: 0.3740--0.5019). Particularly, the reciprocal low-to-high and
high-to-low results indicate that spatial ranking remained informative when
models were transferred between strongly contrasting levels of citywide
epidemic activity.


\begin{table}[htbp]
\centering
\caption{Cross-intensity transportability of the urban structural
susceptibility model. Each structure-only model was trained exclusively using
neighborhood--year observations belonging to one epidemic-intensity group and
evaluated on observations from another group.}
\label{tab:cross_regime}

\resizebox{\textwidth}{!}{
\begin{tabular}{llrrrrrrrr}
\hline
Training & Test & Train $N$ & Test $N$ & RMSE & MAE & $R^2$ &
Pearson & Spearman & Precision@10 \\
\hline
Low & High
& 279 & 372 & 0.9719 & 0.5154 & 0.5019 & 0.7567 & 0.6361 & 0.5 \\

High & Low
& 372 & 279 & 1.5351 & 0.5733 & 0.4620 & 0.7163 & 0.6312 & 0.5 \\

Intermediate & High
& 279 & 372 & 1.0391 & 0.5578 & 0.4306 & 0.7332 & 0.5647 & 0.4 \\

Intermediate & Low
& 279 & 279 & 1.6560 & 0.6957 & 0.3740 & 0.6135 & 0.5279 & 0.3 \\

Low & Intermediate
& 279 & 279 & 1.6669 & 0.6859 & 0.3755 & 0.6140 & 0.6143 & 0.5 \\

High & Intermediate
& 372 & 279 & 1.5842 & 0.6257 & 0.4360 & 0.6927 & 0.5869 & 0.5 \\
\hline
\end{tabular}
}

\end{table}

Because these unrestricted transfer experiments contained unequal numbers of
training observations, an additional matched-sample experiment was performed.
For each held-out year, within-classification and cross-classification training
pools were matched in size before model fitting, and the procedure was repeated
100 times.

After matching training-set size, mean within-classification Spearman
correlation was 0.5756, compared with 0.5399 under cross-classification
training. The transportability gap was therefore 0.0358, with
cross-classification models retaining 93.79\% of the mean ranking performance
obtained from models trained using years in the same epidemic-intensity
classification.

To determine whether this reduction was specifically associated with the
LOW\-/MID/HIGH partition, the matched-sample results were compared with 500
random permutations of the year-level classification labels. Under the
permutation null, the mean transportability gap was 0.0269, with a 95\% null
interval from $-0.0280$ to $0.0937$. The observed gap of 0.0358 was not
unusually large relative to random alternative partitions
(empirical $p=0.8024$). Likewise, the observed performance retention of
93.79\% was consistent with the null distribution
(null mean $=95.47\%$; 95\% null interval: 84.68\%--105.22\%;
empirical $p=0.6228$).

The spatial susceptibility signal therefore remained informative across
contrasting epidemic conditions, but the modest reduction associated with
cross-classification training was not specific to the LOW/MID/HIGH partition.
The evidence supports broad temporal transportability rather than predictive
relationships uniquely organized by these epidemic-intensity groups.


\begin{figure}[htbp]
\centering
\includegraphics[width=0.6\textwidth]
{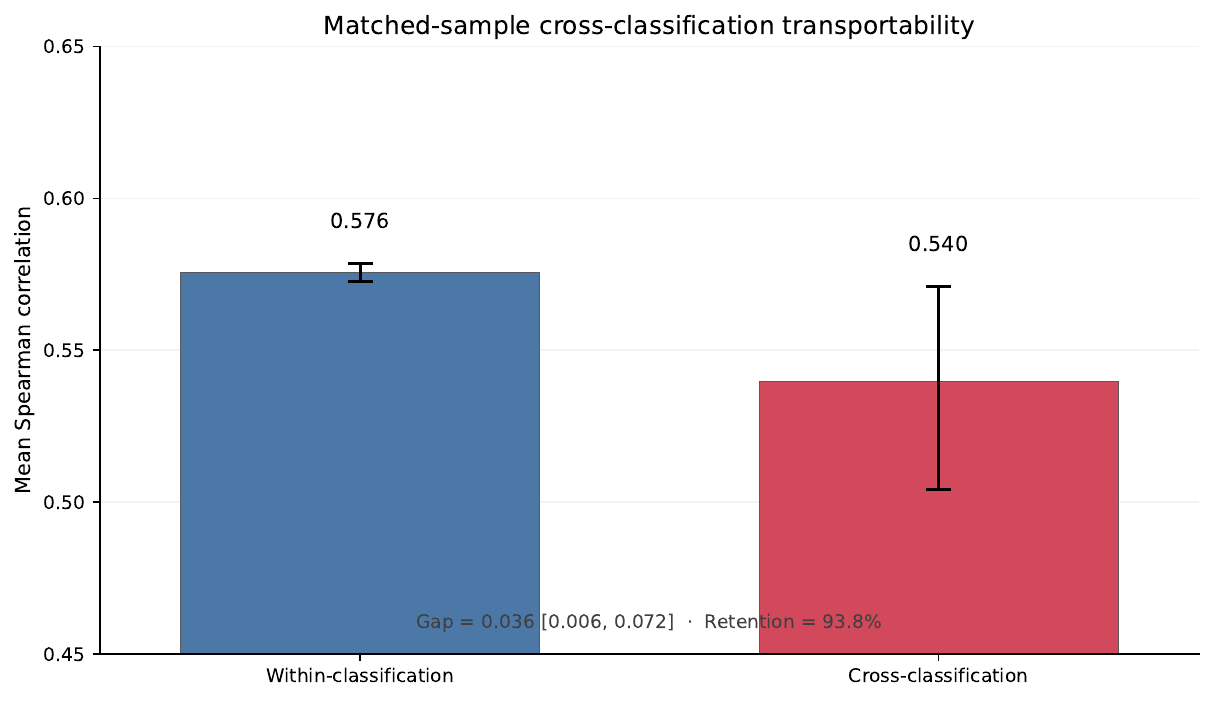}
\caption{Matched-sample evaluation of predictive transportability across
epidemic-intensity classifications. Mean Spearman correlation was 0.5756 for
models trained using observations from the same epidemic-intensity
classification as the held-out year and 0.5399 for models trained using
observations from different classifications. Training sample sizes were
matched before model fitting, and the procedure was repeated 100 times.
The observed transportability gap was 0.0358, corresponding to 93.79\%
performance retention. A permutation analysis based on 500 random
reallocations of year-level classification labels showed that the observed
gap was not unusual relative to alternative partitions of epidemic years
(empirical $p=0.8024$).}
\label{fig:matched_transportability}
\end{figure}



\subsection{Spatial prediction, residual structure, and sensitivity analysis}

We next examined whether temporally out-of-sample predictions reproduced the
persistent geographic organization of dengue susceptibility across Recife.
Observed and predicted susceptibility were converted independently to
within-year percentile ranks and subsequently averaged across the ten study
years for each neighborhood.

The resulting out-of-sample susceptibility surface reproduced major geographic
contrasts in the observed mean annual percentile ranking. Neighborhoods
occupying persistently high or low positions in the observed susceptibility
distribution generally occupied similar positions in the structure-based
predictions. Importantly, this surface does not represent fitted values from a
model trained on the complete epidemiological dataset. Each annual prediction
was generated while the corresponding year was excluded from model fitting,
providing a spatial synthesis of out-of-sample predictions.

Localized discrepancies were represented as predicted minus observed
within-year percentile rank and averaged across years for each neighborhood.
Their spatial organization was then evaluated using Moran's $I$.


\begin{figure*}[htbp]
\centering

\includegraphics[width=\textwidth]
{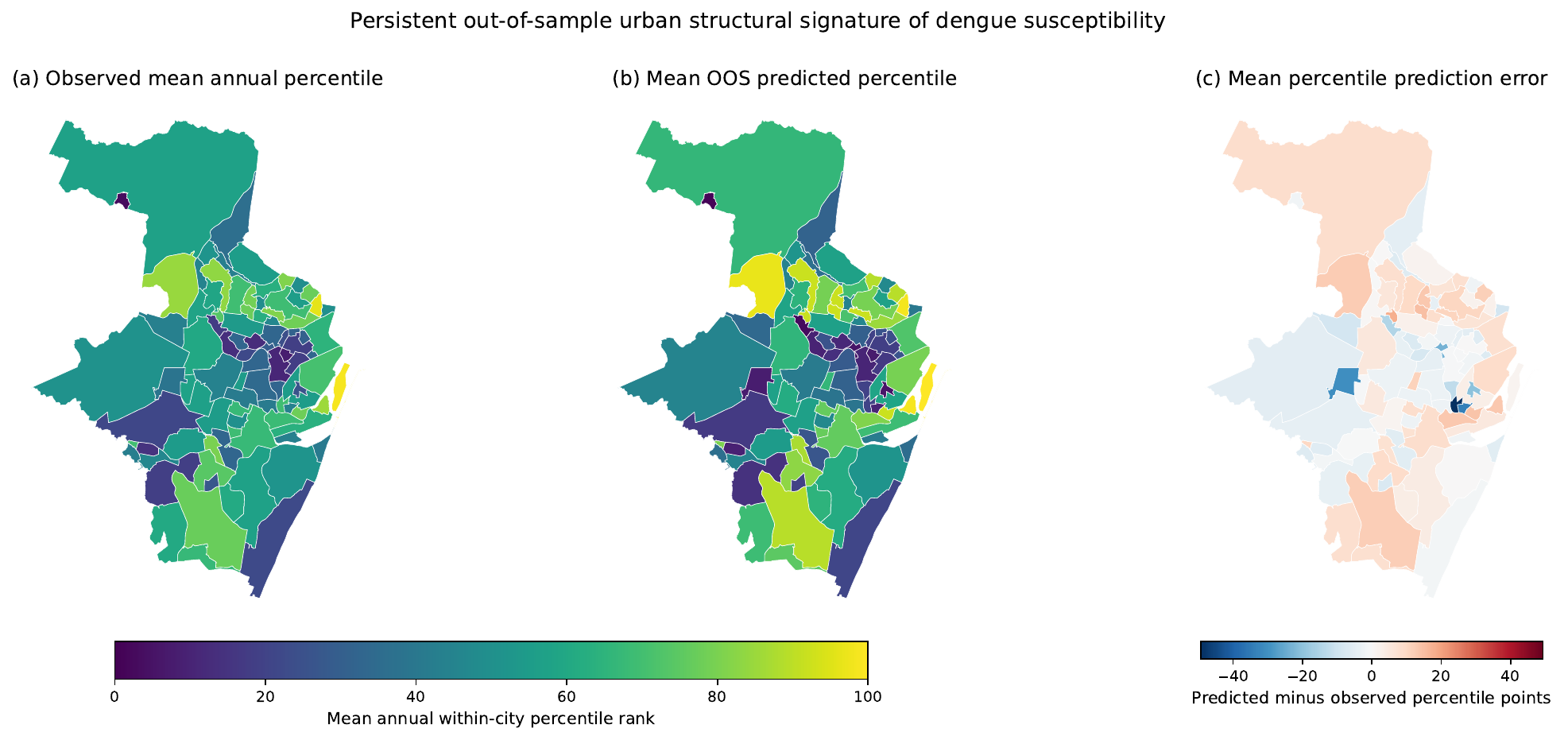}

\caption{Persistent spatial susceptibility signature derived from
leave-one-year-out predictions across Recife neighborhoods.
(A) Mean observed within-city annual percentile rank of dengue susceptibility
over 2015--2024.
(B) Mean out-of-sample predicted annual percentile rank from the
structure-only model, with each annual prediction generated while the
corresponding year was excluded from model training.
(C) Mean percentile prediction error, calculated as predicted minus observed
within-year percentile rank and subsequently averaged across the ten study
years. Panels A and B share the same percentile scale, whereas panel C uses a
diverging scale centered at zero.}

\label{fig:spatial_prediction}
\end{figure*}

Moran's $I$ for the mean annual percentile prediction error was 0.0839,
compared with an expected value of $-0.0109$ under spatial randomness
(simulation $z=1.416$; two-sided permutation $p=0.154$; 999 permutations).
Thus, there was no statistically significant evidence of persistent global
spatial autocorrelation in the mean out-of-sample ranking errors across the
study period.

Year-specific diagnostics nevertheless identified episodic residual spatial
dependence. Positive residual autocorrelation remained statistically
significant after Benjamini--Hochberg correction in 2016
($I=0.1886$), 2020 ($I=0.2079$), 2021 ($I=0.1984$), and 2024
($I=0.1957$), with adjusted $p=0.015$ in each case. The remaining six years
showed no significant residual spatial autocorrelation after correction for
multiple testing. Residual spatial organization was therefore episodic rather
than persistently detectable across the decade.

One particularly large discrepancy was associated with Pau Ferro, a
neighborhood with an unusually small recorded population denominator
(89 inhabitants) and only one dengue case over the 2015--2024 study period.
The principal temporal-generalization and matched-sample transportability
analyses were therefore repeated after excluding this neighborhood.

Exclusion of Pau Ferro reduced the analytical universe from 93 to 92
neighborhoods and from 930 to 920 neighborhood--year observations. Mean LOYO
Spearman correlation changed only modestly, from 0.6393 to 0.6277
($\Delta\rho=-0.0116$). Mean correlations changed from 0.6719 to 0.6615 in
high-incidence years, from 0.6122 to 0.5993 in intermediate-incidence years,
and from 0.6231 to 0.6111 in low-incidence years.

Matched-sample transportability was similarly robust. Mean
within-classification Spearman correlation changed from 0.5756 to 0.5812,
whereas cross-classification correlation increased from 0.5399 to 0.5509.
Consequently, the transportability gap decreased from 0.0358 to 0.0303 and
performance retention increased from 93.79\% to 94.79\%.

Thus, the principal temporal-generalization and transportability findings were
not driven by the atypical Pau Ferro observation.


\begin{table}[htbp]
\centering
\caption{Sensitivity of temporal-generalization and matched-sample
transportability results to exclusion of the spatially atypical Pau Ferro
neighborhood. Differences are calculated as the reduced 92-neighborhood
analysis minus the complete 93-neighborhood analysis.}
\label{tab:sensitivity_pauferro}

\begin{tabular}{lrrr}
\hline
Metric & Full sample & Excluding Pau Ferro & Difference \\
\hline
Number of neighborhoods         & 93      & 92      & -1       \\
Mean LOYO Spearman              & 0.6393  & 0.6277  & -0.0116  \\
High-incidence Spearman         & 0.6719  & 0.6615  & -0.0104  \\
Intermediate-incidence Spearman & 0.6122  & 0.5993  & -0.0129  \\
Low-incidence Spearman          & 0.6231  & 0.6111  & -0.0120  \\
Within-classification Spearman  & 0.5756  & 0.5812  & +0.0056  \\
Cross-classification Spearman   & 0.5399  & 0.5509  & +0.0110  \\
Transportability gap            & 0.0358  & 0.0303  & -0.0055  \\
Performance retention (\%)      & 93.79   & 94.79   & +1.00    \\
\hline
\end{tabular}

\end{table}


\subsection{The persistent predictive pattern is robust to model class}

To assess whether the principal findings depended specifically on the Random
Forest architecture, we repeated the prospective rolling-origin and
leave-one-year-out validation experiments using XGBoost with the same seven
structural predictors and identical temporal train--test partitions. This
analysis was designed as a model-class robustness assessment rather than as a
procedure for selecting the best-performing algorithm.

Under prospective rolling-origin validation, XGBoost achieved a mean Spearman
correlation of 0.5883, compared with 0.5913 for Random Forest. The paired mean
difference was $-0.0030$ (95\% bootstrap CI: $-0.0187$ to $0.0114$;
Wilcoxon $p=0.625$). Under LOYO validation, mean Spearman correlations were
0.6407 for XGBoost and 0.6393 for Random Forest, corresponding to a paired mean
difference of 0.0013 (95\% bootstrap CI: $-0.0072$ to $0.0097$;
Wilcoxon $p=0.846$).

The two model classes also produced highly concordant spatial predictions.
Across the ten LOYO test years, the Spearman correlation between Random Forest
and XGBoost prediction rankings ranged from 0.9856 to 0.9922, with a mean of
0.9896. Agreement between persistent spatial rankings obtained by aggregating
out-of-sample predictions across years was similarly high
($\rho=0.9911$).

Robustness was further evaluated across 24 XGBoost hyperparameter
configurations without selecting configurations according to test-set
performance. Mean LOYO Spearman correlations ranged from 0.6128 to 0.6401,
while prospective performance ranged from 0.5480 to 0.5934. These results
indicate that the principal predictive pattern was not materially dependent on
the Random Forest architecture or on a particular specification of the
alternative tree-based model.

Across all complementary experiments, the evidence consistently supported a
persistent place-based spatial hierarchy of dengue susceptibility. Temporal
generalization collapsed when year-specific spatial correspondence was
destroyed but was reproduced when longitudinal place coherence was preserved,
indicating that persistence cannot be uniquely attributed to the seven measured
structural covariates. Cross-intensity performance remained substantial, while
permutation analyses showed that transportability was not specifically
dependent on the LOW/MID/HIGH partition. Spatial diagnostics and sensitivity
analyses further showed that the principal findings were neither explained by
persistent residual spatial clustering nor driven by the atypical Pau Ferro
neighborhood. Together with the near-identical performance and highly
concordant spatial rankings produced by Random Forest and XGBoost, these results
support measured urban structure as an informative predictive representation
of a broader persistent place-based spatial susceptibility hierarchy.


\begin{figure*}[htbp]
\centering

\includegraphics[width=\textwidth]
{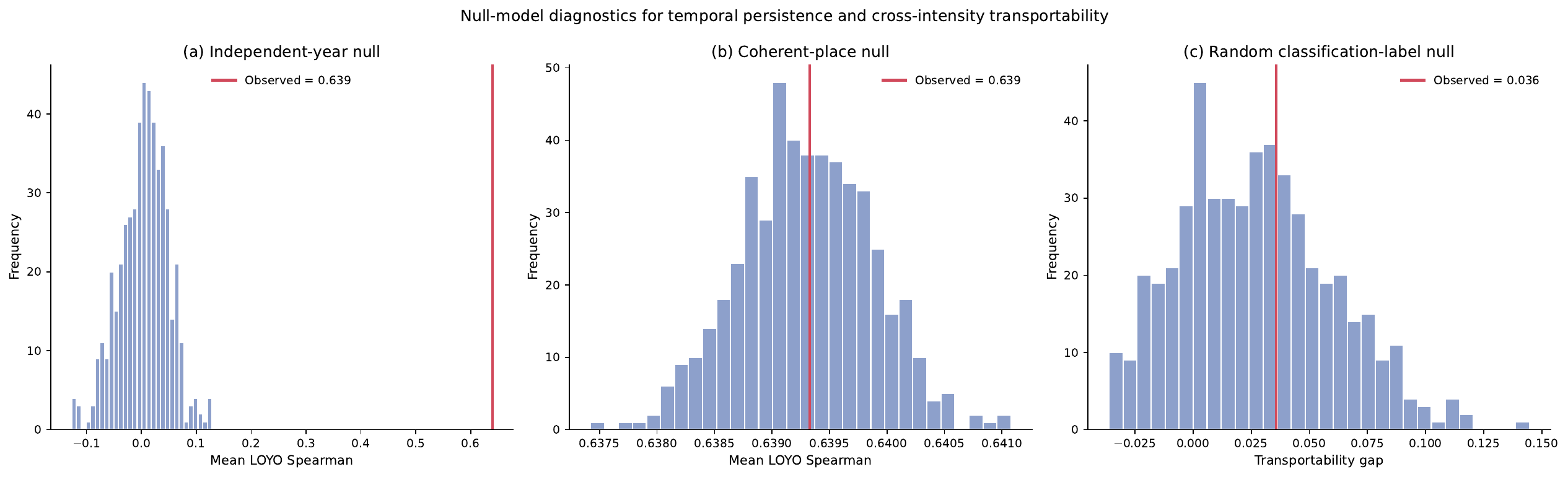}

\caption{Null-model diagnostics for temporal persistence and transportability
of the spatial dengue susceptibility hierarchy.
(A) Distribution of mean LOYO Spearman correlations under independent
year-wise permutation of neighborhood identities, which disrupts persistent
cross-year spatial correspondence.
(B) Distribution under the coherent-place null, which preserves longitudinal
place-level susceptibility trajectories while disrupting their original
correspondence with the measured structural profiles.
(C) Null distribution of the matched-sample transportability gap obtained
from random reallocations of year-level epidemic-intensity classifications.
Vertical lines indicate the corresponding observed statistics. Temporal
generalization collapsed when year-specific spatial correspondence was
disrupted, whereas preserving coherent place-level trajectories reproduced
the observed LOYO performance. The observed transportability gap was not
exceptional relative to alternative partitions of epidemic years.}

\label{fig:null_model_diagnostics}
\end{figure*}

\FloatBarrier
\section{Discussion}

\subsection{Persistent place-based susceptibility and temporal generalization}

The main finding of this study is that the intra-urban distribution of dengue
susceptibility contains a persistent place-based spatial component that remains
predictable across markedly different epidemic years. Across a decade of
pronounced variation in dengue magnitude, models based on relatively stable
urban characteristics consistently recovered substantial correspondence with
the relative spatial ordering of susceptibility across Recife neighborhoods.

This result suggests that the geography of dengue within a city cannot be
understood exclusively as the spatial manifestation of individual epidemic
events. Citywide incidence varied by more than an order of magnitude, yet
neighborhood susceptibility remained partially predictable in years excluded
from model training. Under leave-one-year-out validation, structure-only models
retained positive ranking performance in every held-out year, with a mean
Spearman correlation of 0.6393. Performance ranged from $\rho=0.4907$ in 2020
to $\rho=0.7572$ in 2021. The relevant result is therefore not temporal
invariance, but temporal persistence: the strength of spatial predictability
fluctuated while a substantial place-based signal remained recoverable
throughout the decade.

The permutation experiments substantially refine this interpretation.
Independent year-wise permutation, which destroyed persistent spatial
correspondence across years, reduced mean LOYO performance from an observed
Spearman correlation of 0.6393 to a null mean of 0.0048
($p=0.002$). Conversely, the coherent-place null preserved longitudinal
place-level susceptibility trajectories while disrupting their correspondence
with the original structural profiles and reproduced virtually the entire
observed performance (null mean $=0.6393$; $p=0.485$).

The strongest inference is therefore the existence of persistent place-specific
spatial organization. Persistence itself cannot be uniquely attributed to the
seven measured structural covariates. We use the term \textit{spatial
susceptibility signature} to denote this reproducible place-based hierarchy of
relative dengue susceptibility. Measured urban structure provides an
informative predictive representation of this hierarchy, but not a complete or
causal explanation of its persistence.

The prospective ablation experiments provide complementary evidence regarding
the information represented by these structural characteristics. Structure-only
models achieved higher mean spatial ranking performance than history-only
models, whereas combining structural and epidemiological-history predictors
produced no consistent improvement over the structure-only specification.
Thus, recent dengue occurrence did not fully reproduce the predictive
information contained in the persistent urban indicators considered here.

This does not imply that epidemiological history is unimportant. Previous
dengue occurrence can reflect vector abundance, population immunity,
circulating serotypes, surveillance intensity, and other temporally varying
processes. Rather, structural and historical information operate on partly
different temporal dimensions. The limited gain from their combination may
also reflect partial overlap between the information they encode.

Conceptually, these findings support a distinction between a comparatively
persistent spatial susceptibility landscape and dynamic epidemiological
processes superimposed upon it. Physical, demographic, socioeconomic, and
environmental characteristics of urban space generally evolve more slowly than
annual epidemic intensity. Such persistent characteristics may encode part of
a broader territorial susceptibility landscape upon which time-varying
processes influence the realized magnitude and spatial expression of individual
epidemics.

\subsection{Epidemic intensity and transportability of the spatial signature}

Predictive performance remained substantial during low-, intermediate-, and
high-incidence years. Mean LOYO Spearman correlations were 0.6231, 0.6122,
and 0.6719 for low-, intermediate-, and high-incidence years, respectively,
indicating that spatial predictability did not collapse as citywide dengue
burden changed markedly.

The relative-risk formulation is central to this interpretation. By expressing
neighborhood incidence relative to citywide incidence in the same year, the
target separates, to some extent, annual epidemic magnitude from intra-urban
spatial organization. Widespread amplification of dengue can therefore occur
while part of the relative ordering of neighborhoods remains preserved.
Conversely, low citywide incidence does not imply the disappearance of
persistent spatial contrasts.

Epidemic magnitude and relative spatial susceptibility should therefore be
viewed as related but non-equivalent dimensions of dengue dynamics. Citywide
transmission intensity may respond rapidly to environmental and
epidemiological forcing, whereas the relative geography of susceptibility
contains a more persistent place-based component.

The transfer experiments tested this interpretation more directly by
deliberately separating training and testing according to epidemic intensity.
Spatial ranking remained informative in all six cross-intensity transfers,
including reciprocal low-to-high and high-to-low experiments. After matching
training-set size, mean Spearman correlation was 0.5756 for
within-classification training and 0.5399 for cross-classification training.
Cross-classification models therefore retained 93.79\% of within-classification
ranking performance, corresponding to a transportability gap of 0.0358.

However, the matched permutation experiment showed that this modest reduction
was not specifically associated with the observed LOW/MID/HIGH partition.
The observed gap was compatible with random alternative classifications
($p=0.8024$), as was the observed performance retention
($p=0.6228$). The LOW/MID\-/HIGH grouping should therefore be interpreted as a
descriptive classification of annual epidemic intensity rather than as evidence
of mechanistically or predictively distinct epidemic regimes.

The appropriate interpretation is broad temporal transportability across
contrasting epidemic conditions. This distinction is relevant for public-health
applications because historical surveillance data inevitably combine years
with different epidemiological conditions. The persistence of informative
spatial rankings across such conditions suggests that longer-term
susceptibility mapping need not depend on a single epidemic-intensity context.

\subsection{Spatial, sensitivity, and model-class robustness}

The geographic distribution of out-of-sample predictions provided a spatial
representation of the persistent hierarchy identified through temporal
validation. Aggregated annual LOYO percentile rankings reproduced major
geographic contrasts in observed susceptibility, with neighborhoods that
repeatedly occupied relatively high or low positions generally receiving
similar rankings from the structural models.

Importantly, this surface was constructed entirely from out-of-sample annual
predictions rather than from a model fitted to the complete epidemiological
dataset. It therefore represents a geographic synthesis of recurrent
predictability rather than an in-sample reconstruction of observed dengue
burden.

The spatial organization of prediction errors provides an important
qualification. Mean annual percentile errors showed no statistically
significant persistent global spatial autocorrelation
(Moran's $I=0.0839$, $p=0.154$), suggesting that the model did not
systematically leave the same broad global spatial pattern unexplained
throughout the decade. Residual spatial dependence was nevertheless significant
in 2016, 2020, 2021, and 2024 after correction for multiple testing. Residual
spatial organization was therefore episodic rather than persistently
detectable.

This pattern is compatible with a relatively stable place-based susceptibility
hierarchy overlaid by dynamic spatial processes that become more influential in
particular years. Meteorological conditions, vector abundance, viral
circulation, population immunity, human mobility, surveillance, and
interventions are plausible contributors to such departures. Locations and
periods showing systematic deviations from structural expectations may
therefore be epidemiologically informative rather than simply examples of
prediction failure.

Sensitivity to an extreme small-population spatial unit produced little
evidence that the principal results were driven by denominator instability.
Pau Ferro had a recorded population denominator of only 89 inhabitants and one
recorded dengue case during 2015--2024. Its exclusion reduced mean LOYO
Spearman correlation only modestly, from 0.6393 to 0.6277. In the matched
transportability analysis, within-classification performance changed from
0.5756 to 0.5812 and cross-classification performance from 0.5399 to 0.5509.
The transportability gap consequently decreased from 0.0358 to 0.0303, while
performance retention increased from 93.79\% to 94.79\%. The principal
temporal-generalization and transportability findings were therefore not
driven by this atypical neighborhood.

The close agreement between Random Forest and XGBoost provides an independent
model-class robustness check. Under prospective validation, mean Spearman
correlations were 0.5913 for Random Forest and 0.5883 for XGBoost, while LOYO
performance was 0.6393 and 0.6407, respectively. Neither paired comparison
indicated a systematic difference between model classes.

More importantly, the algorithms generated nearly identical neighborhood
rankings. Mean agreement between Random Forest and XGBoost LOYO prediction
rankings was 0.9896, and the persistent rankings aggregated across years were
correlated at $\rho=0.9911$. Performance also remained substantial across 24
predefined XGBoost hyperparameter configurations without selection according
to held-out performance.

Together, the spatial diagnostics, small-population sensitivity analysis, and
model-class comparison show that the principal predictive pattern is not
readily explained by persistent residual spatial clustering, a single extreme
spatial unit, or the specific Random Forest architecture. These robustness
results strengthen confidence in the reproducibility of the spatial
susceptibility signature, while leaving unchanged the inferential constraint
established by the coherent-place null: robustness of prediction does not
establish that the measured structural variables causally generate the
persistent hierarchy.

\subsection{Implications for sustainable and resilient urban health planning}

The identification of a persistent place-based component of dengue
susceptibility has implications beyond short-term disease prediction. The
structural variables considered here describe dimensions of urban organization,
including population concentration, informal settlement, vegetation,
drainage-related environments, household occupancy, and housing structure.

The results do not establish that these characteristics causally determine
persistent dengue susceptibility. Their predictive value nevertheless suggests
that information relevant to recurrent vulnerability is embedded in the
long-term organization of urban space. Epidemic outbreaks are dynamic
biological and epidemiological phenomena, but they unfold within environments
shaped by infrastructure, land use, housing, demographic concentration, and
territorial inequality.

This distinction separates two complementary forms of public-health
information. Dynamic epidemiological and entomological surveillance identifies
where transmission is occurring or intensifying at a particular moment.
Structural susceptibility mapping addresses a longer temporal horizon,
identifying locations whose persistent urban characteristics are associated
with recurrently elevated relative susceptibility across different epidemic
conditions.

Structural models should therefore complement rather than replace outbreak
forecasting and real-time surveillance. Their potential value lies in
supporting longer-term prioritization of investigation, surveillance capacity,
and urban interventions targeting recurrent vulnerability.

Such information may contribute to decisions concerning drainage and
sanitation, environmental management, vector-control infrastructure,
surveillance allocation, housing interventions, and other place-based
policies. From this perspective, reducing dengue vulnerability is connected
not only to epidemic response but also to the broader objective of building
healthier, more equitable, and more resilient cities.

\subsection{Limitations}

Several limitations should be considered.

First, this is an observational ecological study. Predictive performance and
feature relevance do not establish causal effects, and neighborhood-level
relationships cannot be directly extrapolated to individual dengue risk.

Second, most structural predictors were derived from the 2022 Brazilian
Demographic Census and treated as comparatively persistent throughout
2015--2024. Urban characteristics undoubtedly changed during this period;
the predictors therefore approximate persistent conditions rather than
year-specific exposures.

Third, the coherent-place permutation experiment demonstrates that temporal
persistence cannot be uniquely attributed to the seven measured structural
covariates. Persistent unmeasured neighborhood characteristics may contribute
to the same hierarchy, and the measured variables should be interpreted as a
predictive representation rather than an exhaustive description of its
determinants.

Fourth, dengue surveillance is subject to underreporting, differences in
healthcare access and diagnostic practice, and temporal variation in
notification quality.

Fifth, neighborhood aggregation masks intra-neighborhood heterogeneity and may
introduce scale- and zoning-dependent effects. Very small population
denominators can also destabilize incidence-derived targets, although the Pau
Ferro sensitivity analysis indicated that this issue did not drive the
principal rank-based findings.

Sixth, the structural framework deliberately excludes dynamic determinants
such as meteorological conditions, vector abundance, circulating serotypes,
population immunity, human mobility, and intervention intensity. The objective
was therefore not to reproduce the complete transmission process but to
evaluate persistent spatial information recoverable from urban structural
characteristics.

Seventh, the LOW/MID/HIGH grouping was defined descriptively from annual
citywide incidence and does not identify mechanistically distinct epidemic
states. The permutation experiments provide no evidence that this particular
partition defines uniquely different predictive relationships.

Finally, the analysis establishes temporal transportability within a single
city rather than geographic transportability across urban systems. Recife
provides a demanding longitudinal setting because of its substantial variation
in epidemic magnitude, but external validation is required before the
identified predictive relationships can be generalized geographically.

\subsection{Future research}

A primary next step is to determine which components of the persistent
place-based susceptibility hierarchy are geographically transportable.
Applying the framework to cities with different climatic conditions, urban
morphologies, socioeconomic structures, and dengue histories would help
distinguish city-specific relationships from more general features of urban
arbovirus susceptibility.

A second direction is to decompose persistent place-based susceptibility into
measured and unmeasured components. The coherent-place null indicates that
stable neighborhood-specific heterogeneity contains substantial predictive
information that cannot be uniquely attributed to the seven structural
variables examined here. Additional environmental, infrastructural,
socioeconomic, and mobility-related characteristics may help explain this
persistent organization.

A third direction is to integrate persistent susceptibility with dynamic
epidemic forcing. Meteorological conditions, entomological indicators,
mobility, viral circulation, population immunity, and epidemiological history
could be modeled as time-varying processes acting on a comparatively stable
spatial susceptibility landscape. Such approaches may connect long-term urban
vulnerability with short-term epidemic activation.

Future studies should also investigate alternative spatial scales and explicit
spatiotemporal models where sufficiently reliable data are available.
Fine-scale analyses may reveal heterogeneity concealed by neighborhood
aggregation, while spatial and spatiotemporal models could characterize
episodic residual dependence not represented by the present framework.

Finally, the distinction between persistent place-based susceptibility and
dynamic epidemic intensity may extend beyond dengue. Comparative analyses of
other arboviruses and urban health outcomes could determine whether analogous
spatial hierarchies emerge when persistent territorial conditions interact
with time-varying environmental and epidemiological processes.

\section{Conclusion}

This study provides evidence that the intra-urban distribution of dengue in
Recife contains a persistent place-based spatial hierarchy of susceptibility
that remains detectable across a decade of markedly different epidemic
conditions. Measured urban structural characteristics provided prospective
predictive information beyond recent epidemiological history and recovered
this spatial ordering in complete years excluded from model training.

Null-model experiments clarified both the strength and the limits of this
finding. Temporal generalization depended strongly on persistent place-specific
spatial organization, but this persistence could not be uniquely attributed
to the seven measured structural covariates. Measured urban structure should
therefore be interpreted as an informative predictive representation of a
broader place-based susceptibility hierarchy rather than as its complete or
causal explanation.

This hierarchy remained informative across contrasting levels of citywide
epidemic intensity, while permutation analyses showed that its transportability
was not specifically dependent on the LOW/MID/HIGH classification. Spatial
diagnostics, sensitivity analyses, and replication with XGBoost further
supported the robustness of the principal predictive pattern.

Together, these findings distinguish a comparatively persistent geography of
relative susceptibility from the dynamic processes that determine epidemic
magnitude. Dengue incidence may vary sharply between years while part of the
intra-urban ordering of susceptibility persists. This distinction provides a
useful framework for understanding how long-term territorial conditions and
short-term epidemiological processes jointly shape urban dengue risk.

Structural susceptibility mapping may therefore complement dynamic
epidemiological surveillance by identifying locations where persistent urban
conditions coincide with recurrently elevated relative susceptibility. More
broadly, integrating persistent territorial vulnerability with dynamic
epidemic monitoring may support longer-term prioritization of surveillance,
urban interventions, and strategies for healthier and more resilient cities.

\section*{CRediT authorship contribution statement}

\textbf{Marcílio Ferreira dos Santos:}
Conceptualization, Methodology, Software, Formal analysis, Investigation,
Visualization, Writing -- original draft, Writing -- review \& editing,
Project administration.

\textbf{Cleiton de Lima Ricardo:}
Supervision, Writing -- review \& editing.

\textbf{Andreza dos Santos Rodrigues de Melo:}
Data curation, Writing -- review \& editing.

\textbf{José Dilson Beserra Cavalcanti:}
Funding acquisition, Resources, Writing -- review \& editing.

\textbf{Larissa Suellen Gomes Andrade de Lima Aquino:}
Writing -- review \& editing.

\section*{Funding}

This research benefited from the research infrastructure of the Laboratory of
Intelligent, Sustainable and Educating Territories (LAB-TISE), Universidade
Federal de Pernambuco (UFPE). The establishment and development of LAB-TISE
were supported through parliamentary funding secured under the institutional
coordination of José Dilson Beserra Cavalcanti.

\section*{Acknowledgements}

The authors acknowledge the Laboratory of Intelligent, Sustainable and
Educating Territories (LAB-TISE), Universidade Federal de Pernambuco (UFPE),
for providing the institutional and research environment that supported the
development of this study. The authors also acknowledge the Centro Acadêmico
do Agreste (CAA/UFPE) for institutional support.

\section*{Declaration of competing interest}

The authors declare that they have no known competing financial interests or
personal relationships that could have appeared to influence the work reported
in this paper.

\section*{Data availability}

The dataset used in this study is publicly available through Zenodo as
\textit{Spatial and Socioenvironmental Dengue Dataset of the Recife
Metropolitan Area (2015--2024)} \cite{ferreira2025dataset}, under the DOI
10.5281/zenodo.17364863. The repository contains the spatial and
socioenvironmental data supporting the analyses conducted in this study.
Analytical code used to reproduce the main experiments will be made publicly
available upon publication.

\bibliographystyle{elsarticle-num}
\bibliography{refs}

\clearpage

\appendix


\setcounter{figure}{0}
\setcounter{table}{0}

\renewcommand{\thefigure}{S\arabic{figure}}
\renewcommand{\thetable}{S\arabic{table}}

\makeatletter
\@ifpackageloaded{hyperref}{
    \renewcommand{\theHfigure}{supp.figure.\arabic{figure}}
    \renewcommand{\theHtable}{supp.table.\arabic{table}}
}{}
\makeatother

\section*{Appendix A -- Structural Predictors}

Seven variables were used to characterize relatively persistent demographic,
housing, and environmental characteristics of Recife neighborhoods. Census
variables were derived primarily from the 2022 Brazilian Demographic Census
(IBGE), whereas spatial indicators were obtained from the corresponding
geographic datasets described in the Materials and Methods.

\begin{table}[H]
\centering
\caption{Structural predictors used in the final modeling framework.}
\label{tab:structural_predictors_appendix}

\renewcommand{\arraystretch}{1.3}
\begin{tabular}{p{4.5cm} p{8.5cm}}
\toprule
\textbf{Predictor} & \textbf{Definition} \\
\midrule

Population density &
Population per unit area. \\

Informal-settlement proportion &
Proportion of the spatial unit associated with informal-settlement areas. \\

Vegetation proportion &
Proportion of the spatial unit covered by vegetation. \\

Channel proportion &
Proportion of the spatial unit associated with channels or drainage-related
features. \\

Average residents per household &
Average number of residents per occupied household. \\

Vacant household rate &
Proportion of private households classified as vacant. \\

Collective household rate &
Proportion of housing units classified as collective households. \\

\bottomrule
\end{tabular}
\end{table}

Structural indicators were harmonized to the 93 neighborhoods represented in
the epidemiological panel using official geographic identifiers. All
epidemiological neighborhoods had corresponding geometries in the final
spatial dataset. Queen contiguity among neighborhood polygons produced a
single connected spatial component with no islands. The resulting
row-standardized spatial weights matrix was used in the residual spatial
autocorrelation analyses described in the Materials and Methods.

\section*{Appendix B -- Supplementary Model Performance}

\begin{figure*}[htbp]
\centering
\includegraphics[width=0.95\textwidth]
{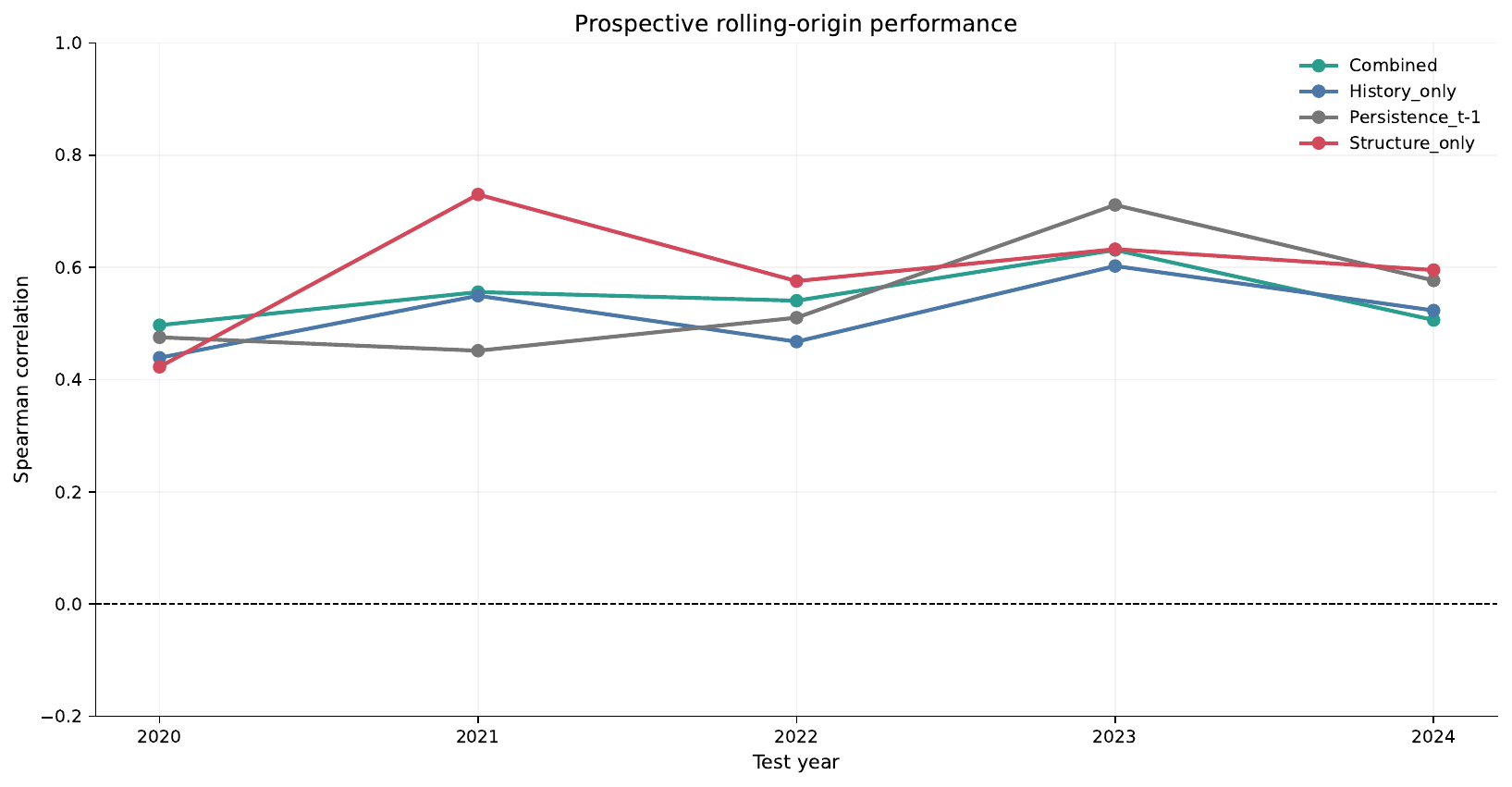}

\caption{Prospective comparison of predictive model specifications.
Performance is shown for the structure-only, history-only, combined, and
persistence-based approaches under temporal out-of-sample validation.
The comparison evaluates the extent to which relatively persistent urban
structural characteristics provide predictive information beyond recent
epidemiological history.}

\label{fig:supp_prospective_models}
\end{figure*}

\begin{figure*}
\centering
\includegraphics[width=0.60\textwidth]
{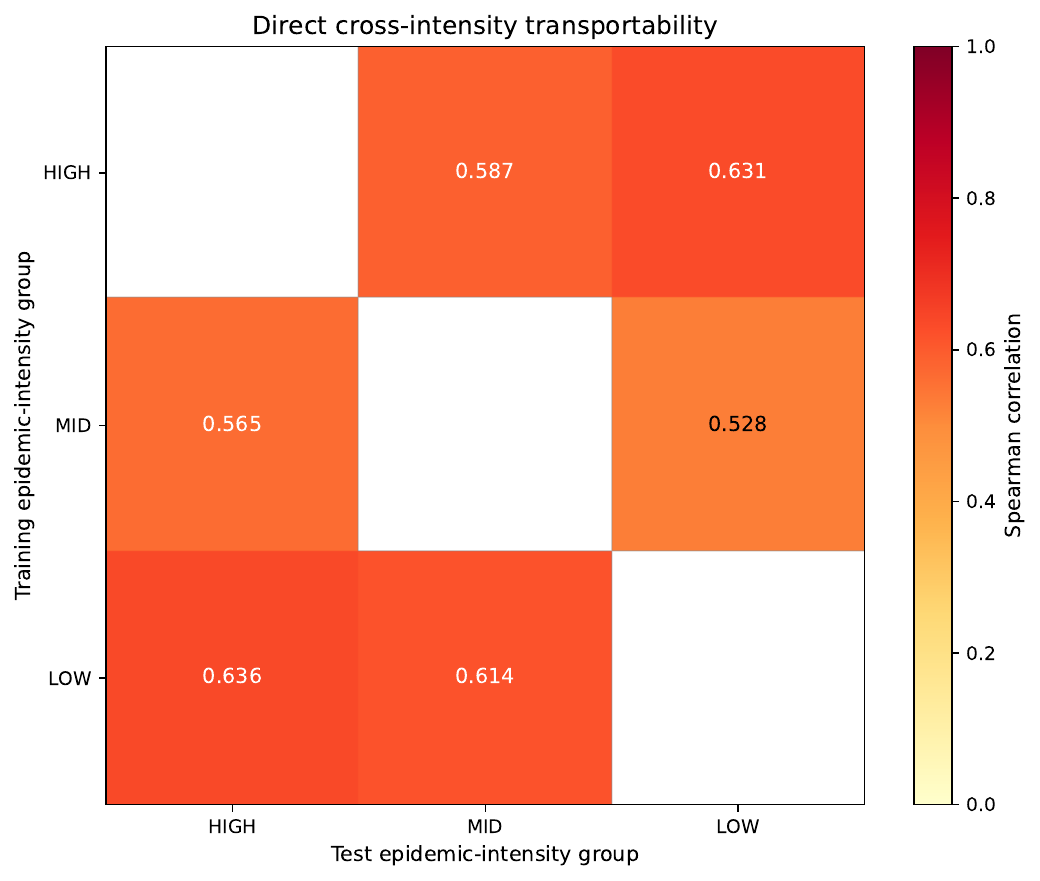}

\caption{Direct transfer of the structure-only model across epidemic-intensity
classifications. Each cell reports the Spearman correlation obtained when the
model was trained using years from one epidemic-intensity classification and
evaluated on years belonging to another classification. These analyses provide
a descriptive assessment of cross-condition transportability and should not
be interpreted as evidence that the LOW, MID, and HIGH classifications
represent intrinsically distinct epidemiological regimes.}

\label{fig:supp_direct_transfer}
\end{figure*}

This appendix provides complementary predictive-performance results supporting
the main temporal-validation analyses. Figure~\ref{fig:supp_prospective_models}
compares the alternative predictor specifications under rolling-origin
validation, whereas Figure~\ref{fig:supp_direct_transfer} summarizes the
direct transfer of the structure-only model across contrasting
epidemic-intensity classifications. These analyses complement the main-text
results by showing both the relative contribution of structural and recent
epidemiological information and the directional consistency of transfer across
different levels of citywide dengue incidence.

\section*{Appendix C -- Supplementary Null-Model Diagnostics}

\begin{figure*}[htbp]
\centering
\includegraphics[width=0.95\textwidth]
{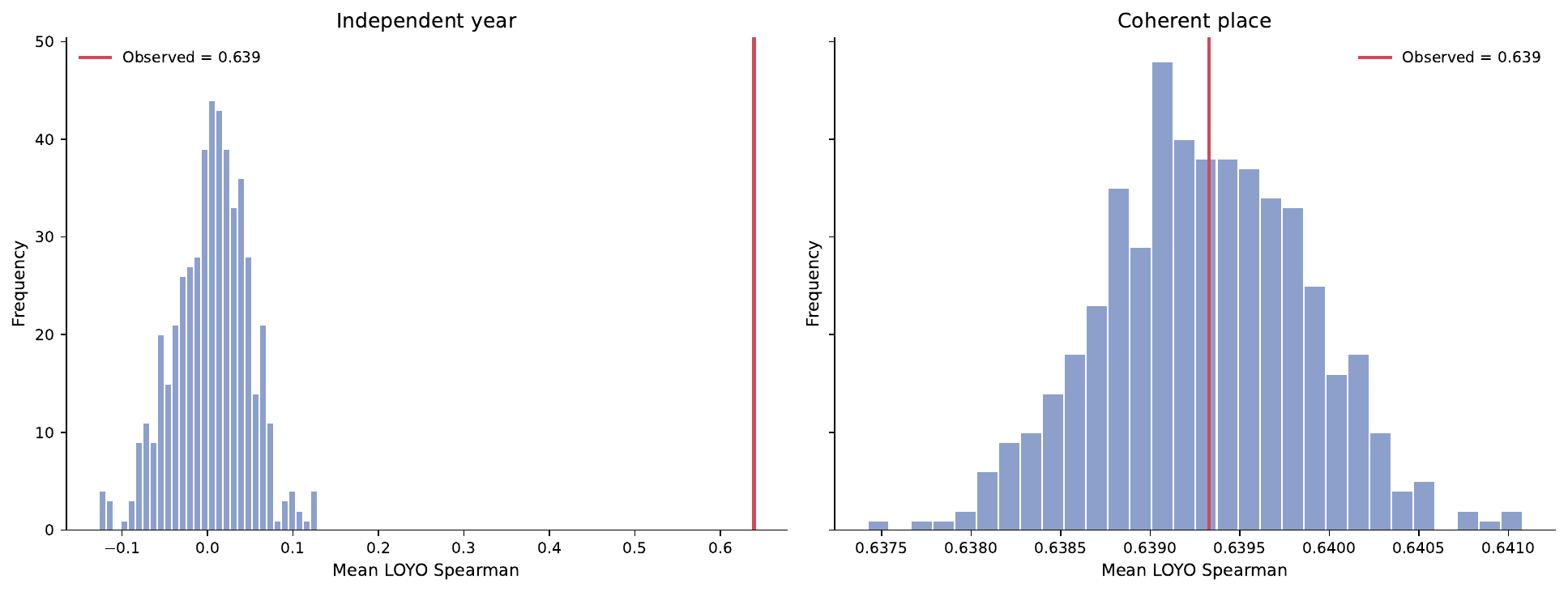}

\caption{Supplementary distributions from the LOYO null-model experiments.
The independent-year null disrupts year-specific spatial correspondence,
whereas the coherent-place null preserves longitudinal place-level coherence
while disrupting its original correspondence with the measured structural
profiles. The observed LOYO performance is shown relative to the corresponding
permutation distributions.}

\label{fig:supp_loyo_nulls}
\end{figure*}

This appendix provides additional visualization of the permutation experiments
used to investigate the source of temporal generalization. The null models were
designed to distinguish predictability associated with persistent
place-specific spatial organization from predictability attributable to the
specific correspondence between the measured structural profiles and observed
dengue susceptibility. Figure~\ref{fig:supp_loyo_nulls} shows the complete
LOYO null distributions underlying the inferential comparisons reported in the
main text.

\FloatBarrier
\section*{Appendix D -- Structural Predictor Importance}

\begin{figure*}[htbp]
\centering
\includegraphics[width=0.90\textwidth]
{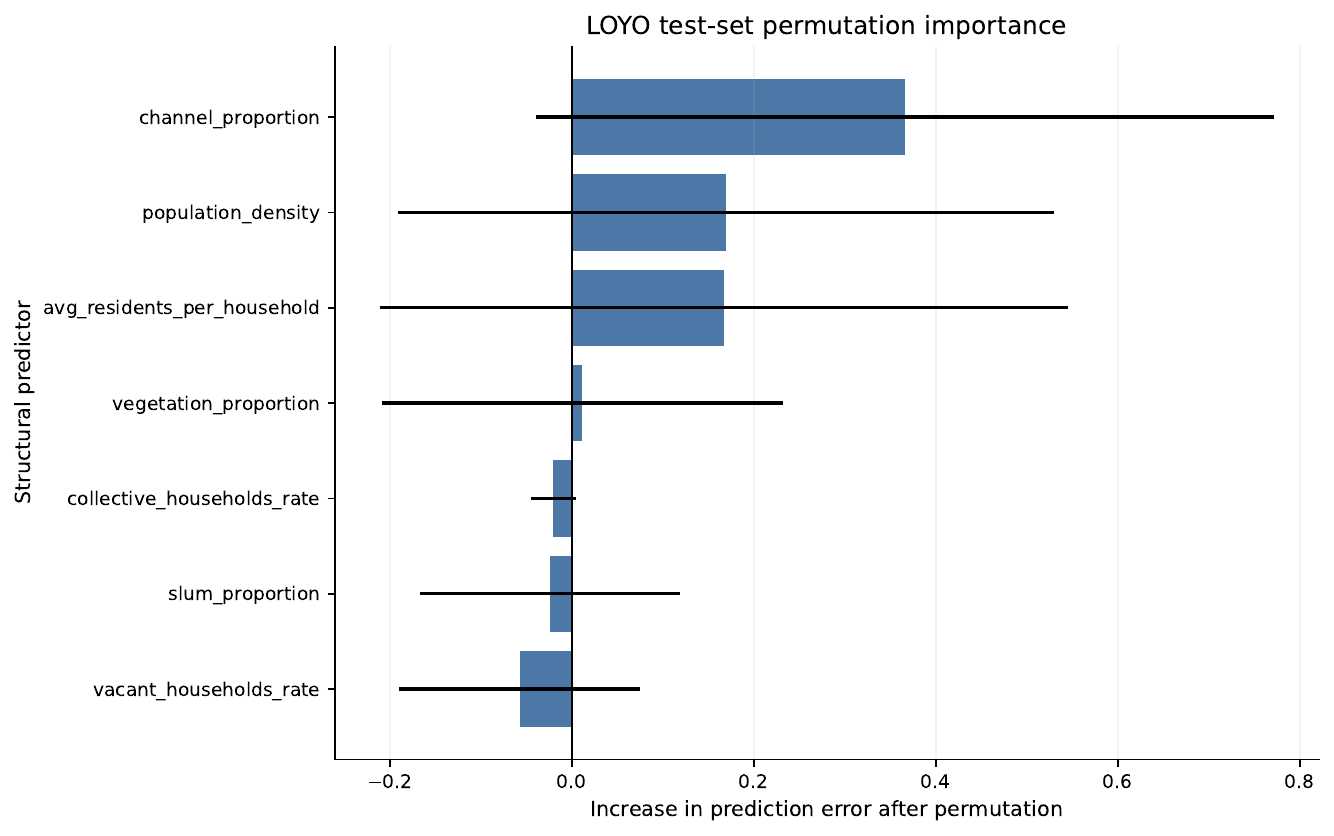}

\caption{Permutation-based predictive importance of the seven structural
predictors in the structure-only modeling framework. Importance values
represent predictive contribution to the fitted model and should not be
interpreted as estimates of causal effects.}

\label{fig:supp_feature_importance}
\end{figure*}

This appendix summarizes the relative predictive contribution of the seven
urban structural indicators used in the structure-only modeling framework.
Because feature-importance measures quantify the contribution of predictors to
model performance rather than causal effects, the results are interpreted as
descriptive evidence of which measured structural characteristics were most
informative for recovering the observed spatial susceptibility hierarchy.
Figure~\ref{fig:supp_feature_importance} presents permutation-based importance
estimates for the structural predictor set.

\end{document}